
\input harvmac.tex
\overfullrule=0pt
\def\half{\textstyle{ { 1\over { 2 } }}}

\def\thalf{\textstyle{{1 \over {2}}}}
\def\tfourth{\textstyle{{1 \over {4}}}}

\def\ts{\thinspace}

\def\ub{\underbar}
\def\myfoot#1#2{{\baselineskip=14.4pt plus 0.3pt\footnote{#1}{#2}}}
%
%
%

\def\CL{{\cal {L}}}

\def\CO{{\cal {O}}}

%
%

%
\def\atc{\alpha_{TC}}

\def\Few{F_\pi}

\def\W+-{W^\pm}
\def\Z0{Z^0}

\def\Ms+-{M^2_\pm}

\def\M+-{M_\pm}

\def\kslash{\raise.15ex\hbox{/}\kern-.57em k}
\def\pslash{\raise.15ex\hbox{/}\kern-.57em p}
%
%
%

\def\gev{{\rm GeV}}
\def\tev{{\rm TeV}}

\def\nb{{\rm nb}}

\def\fb{{\rm fb}}
\def\uhl{10^{34} \ts {\rm cm}^{-2} \ts {\rm s}^{-1}}
\def\ecm{\sqrt{s}}


\def\simge{\mathrel{%
   \rlap{\raise 0.511ex \hbox{$>$}}{\lower 0.511ex \hbox{$\sim$}}}}
\def\simle{\mathrel{
   \rlap{\raise 0.511ex \hbox{$<$}}{\lower 0.511ex \hbox{$\sim$}}}}

\def\slashchar#1{\setbox0=\hbox{$#1$}           
   \dimen0=\wd0                                 
   \setbox1=\hbox{/} \dimen1=\wd1               
   \ifdim\dimen0>\dimen1                        
      \rlap{\hbox to \dimen0{\hfil/\hfil}}      
      #1                                        
   \else                                        
      \rlap{\hbox to \dimen1{\hfil$#1$\hfil}}   
      /                                         
   \fi}                                         %

%

\Title{\vbox{\baselineskip12pt\hbox{BUHEP--94--26}}}
{Technicolor\myfoot{$^{*}$}{Invited talk given at the International
Conference on the History of Original Ideas and \hfill\break
Basic Discoveries in Particle Physics, Erice, Sicily,
29~July--4~August 1994.}}

\centerline{Kenneth Lane\myfoot{$^{\dag }$}{lane@buphyc.bu.edu}}
\smallskip\centerline{Department of Physics, Boston University}
\centerline{590 Commonwealth Avenue, Boston, MA 02215}
\vskip .3in

\centerline{\bf Abstract}
\bigskip

Technicolor, with extended technicolor, is the theory of dynamical
electroweak and flavor symmetry breaking at energies far below the Planck
scale. To motivate it, I describe the most important difficulties of the
standard electroweak model of symmetry breaking by elementary scalar
bosons. I then tell how technicolor deals with these difficulties in a way
that is both technically and physically natural. Finally, I discuss the
problems of technicolor, both past and present.

\bigskip

\Date{10/94}

\vfil\eject

\newsec{INTRODUCTION}

The talks presented so far at this conference have related the history of
most of the great advances of the past half century in particle physics. I
have listened in awe---and in envy---to the stories of the many really
beautiful experiments that form the rock on which our field stands. After
the sad demise of the SSC, hearing these stories has done a great deal to
rekindle my devotion to particle physics. I have also listened with
enormous delight to the stories, in some of which I even had the luck of
playing a small part, recounting our field's theoretical victories. All
that we have heard so far is glorious history---the ghost of Christmas
past.

The theoretical talks you will hear next are about the future. They are
about ongoing attempts to lift the veil that still protects the best kept
secrets of particle physics: the mechanisms of electroweak and flavor
symmetry breaking. Some speakers will review the past---the paths to
``discovery'' and so on. But, make no mistake: at best, the subjects
discussed next have only a future. They'll have a past when and if they
receive experimental verification. Given the declining commitment of
society to science in general, I fear that may not come for some time.

So, despite the conference directors' rquests, my talk about technicolor
will {\ub {not}} be historical. I would love to tell you who did what and
when. But, just now, that is of no interest to anyone but me and them. What
I will do is tell you why I believe in technicolor and extended
technicolor---the theory of {\it dynamical} electroweak and flavor symmetry
breaking; TC and ETC for short. In the second half of my talk, I will tell
you what is wrong with technicolor, both what is often said to be wrong
with it and what I regard to be the real challenges facing technicolor
today.

\newsec{WHY I BELIEVE IN TECHNICOLOR}

\subsec{The Importance of Electroweak and Flavor Symmetry Breaking}

Two great problems face particle physics today. The first is to determine
the origin of electroweak symmetry breaking. Over 30~years ago, Glashow
proposed that electroweak symmetry was $SU(2) \otimes U(1)$, but he
didn't know how to break it
\ref\slg{S.~L.~Glashow, Nucl.~Phys.~{\bf 22} (1961) 579.}.
Making use of the mechanism discovered by Higgs and others
\ref\higgs{P.~W.~Anderson, Phys.~Rev.~{\bf 110} (1958) 827; {\it
ibid.}, {\bf 130} (1963) 439\semi
Y.~Nambu, Phys.~Rev.~{\bf 117} (1959) 648\semi
J.~Schwinger, Phys.~Rev.~{\bf 125} (1962) 397\semi
P.~Higgs, Phys.~Rev.~Lett.~{\bf 12} (1964) 132\semi
F.~Englert and R.~Brout, Phys.~Rev.~Lett.~{\bf 13} (1964) 321\semi
G.~S.~Guralnik, C.~R.~Hagen and T.~W.~B.~Kibble, Phys.~Rev.~Lett.~{\bf
13} (1964) 585.},
Weinberg and Salam produced a model of spontaneously broken $SU(2)\otimes
U(1)$
\ref\sm{S.~Weinberg, Phys.~Rev.~Lett.~{\bf 19} (1967) 1264\semi
A.~Salam, in Proceedings of the 8th Nobel Symposium on Elementary Particle
Theory, Relativistic Groups and Analyticity, edited by N.~Svartholm
(Almquist and Wiksells, Stockholm, 1968), p.~367}
that 't~Hooft and Veltman later showed was a consistent, renormalizable field
theory
\ref\thooft{G.~'t~Hooft, Nucl.~Phys.~{\bf B33} (1971) 173; {\it
ibid}~{\bf B35} (1971) 167; Phys.~Lett.~{\bf 37B} (1971) 195\semi
G.~'t~Hooft and M.~Veltman, Nucl.~Phys.~{\bf
B44} (1972) 189 ; {\it ibid}~{\bf B50} (1972) 318.}.
This model has been confirmed in every important aspect
\ref\aubert{See, in particular, the talks at this conference
by B.~Aubert on the discovery of neutral currents and
P.~Darriulat on the discoveries of the $W$ and $Z$ bosons.},
\ref\pgl{P.~Langacker,{\it Theoretical Study of the Electroweak Interaction
--- Present and Future}, to appear in the  proceedings of the 22$^{\rm nd}$
INS Symposium on Physics with High Energy Colliders, Tokyo, March 1994.}
{\it except one}. No one has found the Higgs boson. And no one
I know believes that the simple one--Higgs--doublet model provides
the correct description of electroweak symmetry breaking. There are many
theoretical proposals, but not a scrap of experimental evidence
indicating what form the electroweak Higgs mechanism takes.

One crucial aspect of electroweak symmetry breaking is known: its
characteristic energy scale $\Lambda_{EW}$ is about $1\,\tev$. This
scale is set by the decay constant of the three Goldstone bosons
transformed via the Higgs mechanism into the longitudinal components,
$W^\pm_L$ and $Z^0_L$, of the weak gauge bosons:
\eqn\weakscale{ \Few \equiv 2^{-\tfourth} G_F^{-\thalf} =
246\,\gev \ts .}
{\ub {\it New physics must occur near this energy scale.}} What form will
it take? The only honest answer theorists can give is that they don't
know. Whatever the new physics, it was the energy scale of
$1\,\tev$ and the size of typical QCD and electroweak cross sections at this
energy, $\sigma \simeq 1\,\nb$--$1\,\fb$, that determined the energy and
luminosity requirements of the Superconducting Super Collider: $\ecm =
40\,\tev$ and $\CL = 10^{33}$--$\uhl$
\ref\ehlq{E.~Eichten, I.~Hinchliffe, K.~Lane and C.~Quigg,
Rev.~Mod.~Phys.~{\bf 56} (1984) 579.}.
Now the task of the SSC must be taken up by CERN's Large Hadron Collider
and the experiments being prepared for it. I wish the LHC godspeed.

The second great mystery confronting particle physics is the origin of
flavor symmetry and its breaking. This problem has been with us for over 50
years and is epitomized by Rabi's famous question about the muon: ``Who
ordered that?'' (I have a distinct vision of Rabi sitting at lunch at the
Moon Palace on Broadway asking that question, albeit in a different
context.) Nowadays we ask: What is the origin of quark and lepton flavors?
Why do they come in three identical generations? What is the origin of
quark and lepton masses and mixings? Why is $m_\nu \cong 0$? ($\equiv 0$??)
Why is $m_t = 174\,\gev$? ($\equiv 2^{-3/4}\ts G_F^{-1/2}$??)
\ref\cdfpr{F.~Abe, et al., The CDF Collaboration, Phys.~Rev.~Lett.~{\bf
73} (1994) 225; Phys.~Rev.~{\bf D50} (1994) 2966.}
What causes CP~violation? If there is any doubt of the importance of flavor
physics {\it beyond} the first generation, just look at the number of talks
on it at this conference:~16
\ref\flavor{They were given by by O.~Piccione, R.~Dalitz, R.~Adair,
A.~Zichichi,
M.~Perl, S.~Ting, S.~Glashow, A.~de~Rujula, G.~Goldhaber, D.~Kaplan,
V.~Telegdi, R.~Turlay, C.~Jarlskog, N.~Samios, F.~Boehm, and B.~Sadoulet.}.

Flavor physics is difficult, harder by far to explain than electroweak
symmetry breaking. Two things make it so hard: first, unlike the case of
electroweak symmetry breaking, the energy scale of flavor symmetry is
unknown. It is probably as high or higher than the weak scale,
$\Lambda_{EW} \simeq 1\,\tev$, but that isn't much help. Second,
low--energy physics (where we are now) is essentially
flavor--conserving---in the sense of the GIM mechanism
\ref\gim{S.~L.~Glashow, J.~Iliopoulos and L.~Maiani, Phys.~Rev.~{\bf D2}
(1970) 1285.}.
The mechanism by which flavor symmetry is broken is hidden from us in the
most fiendishly clever way. It has withstood the concerted efforts of the
past 20~years to pry it out.

Is the answer to Rabi's question to be found in accessible physics---in
experiments performed at the TeV~energy scale---or in the inaccessible
mists of Planck scale physics? Are we going to attack the flavor question
now, experimentally as well as theoretically, or are we going to give up?
Do we theorists tell our experimental colleagues not to bother looking,
that flavor physics is too far away? Yes, there are 16~talks at this
conference on flavor physics, but 13 of them deal with experiments of 20 or
more years ago. Three deal with important ongoing searches, but {\it none}
of those are at the highest energy machines built or talked about.
High--energy colliders are where the most direct searches for flavor
physics have occurred. It would be good to hear more of the plans for
tackling flavor physics at these machines.

\subsec{Difficulties of Elementary Higgs Boson Models}

I said above that no one I know seriously believes in
the standard one--doublet Higgs model. I think this statement applies to
{\it all} elementary Higgs models {\it except} those employing
supersymmetry
\ref\susy{For reviews of supersymmetry and its phenomenology, see
H.~E.~Haber and G.~L.~Kane, Phys.~Rept.~{\bf 117} (1985) 75\semi
S.~Dawson, E.~Eichten and C.~Quigg, Phys.~Rev.~{\bf D31} (1985) 1581.}.
Now I'll tell you why that is for nonsupersymmetric models. I'll mention
SUSY later on.

Elementary Higgs boson models provide no explanation of why electroweak
symmetry breaking occurs and why it has the scale $\Few$. The Higgs doublet
self--interaction potential is
\eqn\vphi{V(\phi) = \lambda\ts (\phi^\dagger \phi - v^2/2)^2 \ts,}
where $v$ is the vacuum expectation of the Higgs field $\phi$ {\it when}
$v^2 \ge 0$. But what dynamics makes $v^2 > 0$? Where does the value $v
= \Few = 246\,\gev$ come from? This is to be compared with QCD, where we
understand the magnitude of {\it all} masses in terms of the scale
$\Lambda_{QCD}$ at which the gauge coupling $\alpha_S$ becomes strong.

Elementary Higgs boson models are {\it unnatural}. The Higgs boson's mass,
$M_H = \sqrt{2 \lambda} v$, and the vacuum expectation value itself are
{\it quadratically} unstable against radiative corrections. Thus, there is
no reason why these two parameters should be much less than the
energy scale at which the essential physics of the model changes, e.g., a
unification scale ($M_{GUT} \simeq 10^{16}\,\gev$?) or the Planck scale
($M_P \simeq 10^{19}\,\gev$)---unless an incredibly fine tuning of
parameters to one part in, say, $M_P^2/M_H^2 \sim 10^{34}$ occurs
\ref\natural{K.~G.~Wilson, unpublished; quoted in L.~Susskind,
Phys.~Rev.~{\bf D20} (1979) 2619\semi
G.~'t~Hooft, in {\it Recent Developments in Gauge Theories}, edited by
G.~'t~Hooft, et al. (Plenum, New York, 1980).}.

Another serious problem of elementary Higgs boson models is that they are
{\it ``trivial''}
\ref\trivial{See, for
example, R. Dashen and H. Neuberger,
Phys.~Rev.~Lett.~{\bf 50} (1983) 1897\semi J.~Kuti, L.~Lin and Y.~Shen,
Phys.~Rev.~Lett.~{\bf 61} (1988) 678\semi A.~Hasenfratz, et
al.~Phys.~Lett.{\bf 199B} (1987) 531\semi G.~Bhanot and K.~Bitar,
Phys.~Rev.~Lett.~{\bf 61} (1988) 798.}.
This is the disease of QED diagnosed 40~years ago by Landau and Pomeranchuk
\ref\landau{L.~D.~Landau and I.~Pomeranchuk, Dokl.~Akad.~Nauk~{\bf 102}
(1955) 489.}
and stressed here in D.~Gross's talk. In the minimal one--doublet model,
the Higgs boson interaction strength $\lambda(M)$, at the energy scale $M$,
is given to a good approximation by
\eqn\lamtriv{
\lambda(M) \cong {\lambda(\Lambda) \over {1 + (24 /16 \pi^2)\ts
\lambda(\Lambda) \ts \log (\Lambda /M)}} \ts\ts.}
This vanishes for all $M$ as the cutoff $\Lambda$ is taken to infinity, so
that the Higgs boson is noninteracting, hence the theory is said to be
``trivial''. This feature has been shown to be true in a general class of
two--Higgs doublet models
\ref\dk{R.~S.~Chivukula and D.~Kominis, Phys.~Lett.~{\bf 304B} (1993) 152.},
and may be true of all Higgs models.

The import of triviality is that elementary--Higgs Lagrangians must be
considered to describe {\it effective} theories. They are meaningful only for
scales $M$ below some cutoff $\Lambda_\infty$ at which new physics sets in.
The larger the Higgs couplings are, the lower the scale $\Lambda_\infty$.
This relationship translates into the so--called triviality bounds on Higgs
masses. For the minimal model, the connection between $M_H$ and
$\Lambda_\infty$ is
\eqn\triv{
M_H(\Lambda_\infty) \cong \sqrt{2 \lambda(M_H)} \ts v = {2 \pi v \over
{\sqrt{3 \log (\Lambda_\infty/M_H)}}} \ts.}
The Higgs mass has to be somewhat less than the cutoff in order for the
effective theory to have {\it some} range of validity. From lattice--based
arguments~\trivial, $\Lambda_\infty \simge 2 \pi M_H$. Since $v$ is fixed
at 246~GeV in the minimal model, this implies the triviality bound $M_H
\simle 700\,\gev$. If the standard Higgs boson were to be found with a mass
this large or larger, we would know for sure that additional new physics is
lurking in the range of just a few~TeV\foot{Triviality is not much of an
issue for supersymmetric theories. The Higgs self--couplings, and the Higgs
boson masses, are so small that the cutoff approaches $M_P$}.

Finally, elementary Higgs models tell us nothing about flavor.
No light is shed on why there are color--singlet integrally--charged leptons
and color--triplet fractionally--charged quarks, nor why there are
three generations of them (though each generation is an {\it
anomaly--free} combination
\ref\anomaly{C.~Bouchiat, J.~Iliopoulos and Ph.~Meyer, Phys.~Lett.~{\bf
38B} (1972) 519.}).
The flavor--symmetry breaking Yukawa couplings of the Higgs boson to
fermions are arbitrary free parameters, ranging from zero for neutrinos to
one for the top quark. As far as we know, and as string theorists would
have us believe, it is a logically possible state of affairs that we will
not understand flavor until we understand the physics of the Planck scale.
To me, that is a philosophically unsatisfying state of affairs. (Nor
can I believe in a desert with no new interactions between the weak scale
and the GUT or Planck scale. ``Nothing new'' has never happened before in
physics and there is no reason that it should start to happen at the weak
scale.).

\subsec{Dynamical Electroweak and Flavor Symmetry Breaking}

Technicolor and extended technicolor
\ref\kltasi{For a recent review, see K.~Lane, {\it An Introduction to
Technicolor}, Lectures given June~30--July~2 1993 at the Theoretical
Advanced Studies Institute, University of Colorado, Boulder, published in
``The Building Blocks of Creation'', edited by S.~Raby and T.~Walker, World
Scientific (1994).}
are the principal attempts to address these issues
in a framework that we know, a theory whose fundamental constituents are
just gauge bosons and fermions. Although such a theory has {\it no
elementary scalars}, it can exhibit spontaneous chiral symmetry breaking
\ref\nambu{Y.~Nambu and G.~Jona--Lasinio, Phys.~Rev.~{\bf 122}
(1961) 345.},
\ref\goldstone{J.~Goldstone, Nuovo Cimento~{\bf 19A} (1961) 154\semi
J.~Goldstone, A.~Salam and S.~Weinberg, Phys.~Rev.~{\bf 127} (1962) 965.}.
The working example is QCD.

Imagine that there is a new, asymptotically free gauge interaction,
called ``technicolor'', with gauge group $G_{TC}$, and gauge coupling
$\atc$ that becomes strong in the vicinity of a few hundred~GeV
\ref\swtc{S.~Weinberg, Phys.~Rev.~{\bf D19} (1979) 1277.},
\ref\lstc{L.~Susskind, Phys.~Rev.~{\bf D20} (1979) 2619.}.
In simple technicolor models, $G_{TC} = SU(N_{TC})$ and there are $N_D$
doublets of left-- and right--handed technifermions, $T_{iL,R} = (U_i,
D_i)_{L,R}$. These belong to equivalent complex irreducible representations
of this gauge group. If, just as for quarks and leptons, the $T_L$ are
assigned to electroweak $SU(2)$ as doublets and the $T_R$ as singlets, they
necessarily are {\it massless} and their strong technicolor interactions
are invariant under the chiral flavor symmetry
\eqn\tgchi{G_\chi = SU(2 N_D)_L \otimes SU(2 N_D)_R \supset SU(2)_L \otimes
SU(2)_R \ts.}
The technifermion part of the electroweak group $SU(2) \otimes U(1)$ is
contained in this $SU(2)_L \otimes SU(2)_R$.

We know from QCD that, when $\atc$ becomes strong, this chiral symmetry
breaks spontaneously to the diagonal subgroup $SU(2 N_D)_V \supset
SU(2)_V$. There result $4 N_D^2 - 1$ massless Goldstone bosons. Three
of these are absorbed as the longitudinal components
of the $W^\pm$ and $Z^0$ weak bosons---this is the dynamical Higgs
mechanism
\ref\dynhiggs{R.~Jackiw and K.~Johnson, {\it Phys.~Rev.}~{\bf D8}
(1973) 2386\semi
J.~Cornwall and R.~Norton, {\it Phys.~Rev.}~{\bf D8} (1973) 3338\semi
E.~Eichten and F.~Feinberg, {\it Phys.~Rev.}~{\bf D10} (1974) 3254.}
---and they acquire the masses
\eqn\mwztc{M_W = \half g \Few \ts, \qquad
M_Z = \half \sqrt{g^2 + g'^2} \Few = M_W/\cos\theta_W \ts.}
Here, $g$ and $g'$ are the $SU(2)$ and $U(1)$ couplings and $\tan\theta_W =
g'/g$. The relation $M_W = M_Z \cos\theta_W$, experimentally verified to
better than one percent, is a consequence of the remnant ``custodial''
$SU(2)_V$ symmetry
\ref\marvin{M.~Weinstein, Phys.~Rev.~{\bf D8} (1973) 2511.},
\swtc, \lstc.
Thus, technicolor neatly describes the dynamics of electroweak symmetry
breaking: it is the familiar phenomenon of chiral symmetry breaking that
occurs in QCD.

Technicolor is an asymptotically free gauge interaction, with $\atc$ rising
slowly from a small value at the Planck scale (or, perhaps, some
very--grand unification scale) to a value of order one at the lower scale
$\Lambda_{TC}$. As in QCD, it is then {\it natural} that $\Lambda_{TC}$,
$\Few$ and technihadron masses are all of the {\it same} order of magnitude
and very much smaller than $M_P$. The known value $\Few \simeq 250\,\gev$
then tells us that $\Lambda_{TC} \simeq 1\,\tev$. These mass scales are
stable under renormalization and no fine--tuning of parameters is required.
(In any technicolor theory, as in QCD, there will be a rich spectrum of
technihadrons with masses of order $\Lambda_{TC}$. The details of this new
phenomenology will depend on the group structure and technifermion content
of the model; see, for example, Refs.~\ehlq\ and \ref\multi{K.~Lane and
E.~Eichten, Phys.~Lett.~{\bf 222B} (1989)~274 \semi K.~Lane and
M~V.~Ramana, Phys.~Rev.~{\bf D44} (1991)~2678.}.) On a logarithmic scale
appropriate to running gauge couplings, this is not much larger than the
QCD scale of $1\,\gev$ and, so, $1\,\tev$ is quite plausible from this
point of view. Asymptotic freedom also means that there is no Landau pole
preventing the infinite--cutoff limit in the running coupling $\atc$; that
is, technicolor is {\it nontrivial}.

Technicolor, by itself, does not address the questions of flavor. In fact,
quarks and leptons are still massless because TC does not communicate
electroweak symmetry breaking to them. To solve this problem in the spirit
of technicolor, i.e., naturally, without the introduction of elementary
scalars, it is necessary to invoke still more gauge interactions that
explicitly break the chiral symmetries of quarks and leptons. The most
obvious and economical way to do this is to combine quarks, leptons and
technifermions into the {\it same} representations of an enlarged gauge
group, called extended technicolor
\ref\sdlsetc{S.~Dimopoulous and L.~Susskind, Nucl.~Phys.~{\bf B155}
(1979) 237.},
\ref\eekletc{E.~Eichten and K.~Lane, Phys.~Lett.~{\bf 90B} (1980) 125.}.

At very high energies (but far below $M_{GUT}$ and $M_P$),
quarks, leptons, and technifermions are {\it unified}.  Then, at one or
perhaps a sequence of high scales, $\Lambda_{ETC} \gg \Lambda_{TC}$, the
ETC gauge symmetry is broken down to TC. Quarks and leptons are those
fermions without residual (unbroken) TC interactions. In this way, the
number and types of quarks and leptons are determined by the ETC
representations to which they belong. The broken ETC interactions linking
technifermions to quarks and leptons break the latter's chiral symmetries
and give them mass {\it when} technifermions acquire their own dynamical
masses. Roughly speaking, quark and lepton masses are given by
\eqn\qmass{m_{q,\ell} \approx {\Lambda^3_{TC} \over {\Lambda^2_{ETC} \ts.}}}
Thus, extended technicolor, constructed without elementary scalars,
is a {\it dynamical, natural} explanation of flavor symmetry
and its breaking. We shall see in the next section that, for the light
quarks and leptons, $\Lambda_{ETC}$ needs to be 100s of TeV.

Extended technicolor, like all attempts to explain the physics of flavor,
has so far resisted a simple and attractive implementation. We do not have
the ``standard ETC model'' yet. This, in my opinion, is the most important
thing wrong with technicolor. But, let's also see what others say.

\newsec{WHAT'S WRONG WITH TECHNICOLOR?}

Technicolor is not a popular subject these days. The SPIRES HEP database
can be used to find how many papers have been written on a subject in the
past 20~years by giving it one or more title keywords. On September 28, I
found the following frequencies\foot{There is very little double--counting
in this list. No paper has both ``supersymmetry'' and ``supersymmetric'' in
the title. There are 55 titles with supersymmetric and string; 56 have
supersymmetry and string; only 200 have cosmic and string.}:

\itemitem{} Technicolor---233

\itemitem{} Technipion---15

\itemitem{} Technipions---4

\itemitem{} Walking (Technicolor)---15

\itemitem{} Techni---1

\medskip

\itemitem{} Supersymmetric---2869

\itemitem{} Supersymmetry---2059

\itemitem{} SUSY---354

\itemitem{} MSSM---31

\itemitem{} String---4458

\itemitem{} Superstring---1284

\itemitem{} Super String---7 (including ``Full Power Test of a {\underbar
{String}} of Magnets Comprising a Half Cell of the Superconducting
{\underbar {Super}} Collider'')

\medskip

\noindent This is a 40:1 ratio in favor of supersymmetry.  When I looked
for experimental ``search'' papers, I found 71 on supersymmetry, three on
technicolor. These statistics are harder to gather in a hurry than the
theory ones, so my ratio may be off a bit here. This conference has one
talk on technicolor, five talks on superwhatever. I am glad that the
organizers did not allocate talks according to their frequency in the
SPIRES listings.

What is the basis for this overwhelming vote of confidence in
supersymmetry? It can't be experimental support; there isn't
any\foot{SUSY enthusiasts tout the apparent unification of the
$SU(3)\otimes SU(2)\otimes U(1)$ couplings at the scale $M_X \simeq
10^{16}\,\gev$; see S.~Dimopoulos' talk at this conference. I am not
convinced that even this happens. There is enough uncertainty in $\alpha_S$
to throw unification of the couplings at a single $M_X$ into doubt.}. Some
say it is because supersymmetry is so beautiful. Many of these people point
to its connection with gravity---superstrings are the only known consistent
quantum theories that include gravity. Some say that supersymmetry is the
only path open beyond the standard model. In this, they are at least
implicitly comparing supersymmetry to technicolor\foot{Or its variant
involving {\it composite} quarks and leptons; see
Ref.~\ref\comp{G.~'t~Hooft, in {\it Recent Developments in Gauge Theories},
edited by G.~'t~Hooft, et al. (Plenum, New York, 1980).}} and basing their
judgement on the perceived failures of the latter. Since my talk
is about technicolor, I will concentrate on examining these comparisons.
In preparing this section, I have made much use of Haber's lists of the
``successes'' of and ``challenges'' to supersymmetry
\ref\haber{H.~E.~Haber, ``The Supersymmetric Top--Ten Lists'', invited
talk presented at the Workshop on Recent Advances in the Superworld,
Houston Advanced Research Center, April 14--16, 1993.}.

\bigskip

\item{} {\sl Technicolor is not unique in providing a natural,
dynamical explanation for electroweak symmetry breaking and for stabilizing
Higgs masses and vacuum expectation values well below the Planck scale.
Supersymmetry does these things also.}

\medskip

\noindent Here's how it works \haber: We start with the observation that,
since the world is not supersymmetric, supersymmetry must be broken. The
favored mechanism for this is known as the ``hidden sector scenario''. In
this scenario, supersymmetry is broken at some high mass scale,
$\Lambda_{SUSY}$, but in a particle sector that communicates {\it only
through gravity} with the ``visible sector'' of all known particles and
their superpartners. In the visible sector, then, the {\it effective} scale
of supersymmetry breaking, called $M_{SUSY}$, is suppressed by powers of
the Planck scale, $10^{19}\,\gev$: $M_{SUSY} \simeq \Lambda^n_{SUSY} /
M_P^{n-1}$. To stabilize electroweak symmetry breaking at $\Lambda_{EW}
\simeq 1\,\tev$, we shall see that $M_{SUSY}$ must be order $\Lambda_{EW}$.
In popular models, $n = 2$~or~3, so that $\Lambda_{SUSY} \simge
10^{11}\,\gev$ generates the desired $M_{SUSY}$. SUSY is vague about where
$\Lambda_{SUSY}$ comes from.

\noindent Now, in a supersymmetric theory, a chiral symmetry that keeps
fermions massless will keep their scalar partners massless as well. Once
supersymmetry is broken, these scalars can acquire a mass, but this mass is
logarithmically (not quadratically) renormalized and it is at most of order
the effective breaking scale, $M_{SUSY}$. This mechanism is invoked for the
Higgs supermultiplets (two are required in minimal models). That's how
Higgs masses can be stabilized at values well below $M_P$.

\noindent Finally, here is how electroweak symmetry breaking occurs: If all
couplings are relatively weak, then renormalization effects are mild and
particle masses renormalized at the Planck scale will not be very different
from those at the weak scale. Then, in some supergravity models, the
following miracle happens: Suppose the squared Higgs masses at the Planck
scale are all positive and of order $(100\,\gev)^2$. When the
renormalization equations for these masses are used to evolve them down to
low energy, the large top--quark Yukawa coupling drives one of the $M_H^2$
negative. Thus, a Higgs vacuum expectation value forms and electroweak
symmetry breaks. {\it If} $\Lambda_{SUSY}$ has a value that gives
$M_{SUSY} = \CO(1\,\tev)$ and {\it if} all other mass parameters in the
superpotential are comparable, then the Higgs vacuum expectation values
are stabilized below $1\,\tev$.

\noindent Those last two conditions are the rub: First, the origin of
$\Lambda_{SUSY}$ is unclear and its magnitude uncertain. Furthermore, the
superpotential contains a term, the so-called $\mu$-term, that is required
to avoid a massless axion. The mass parameter $\mu$ enters the Higgs
bosons' $M_H^2$--matrix. There is no reason that $\mu$ is much less than
$M_P$. To maintain electroweak symmetry breaking at $1\,\tev$, $\mu$ is
chosen to be $\CO(M_{SUSY}) = \CO(\Lambda_{EW})$. By the nonrenormalization
property of supersymmetry, {\it any} value of $\mu$ is radiatively stable.
Thus, $\Lambda_{EW} \ll M_P$ is natural, but only in a technical sense: a
fine--tuning of $\mu$ is required in lowest--order perturbation theory in
order that $\mu \sim \Lambda_{EW} \ll M_P$. Once this value is chosen, it
is stable under renormalization; supersymmetry ``sets it and forgets it''.

\bigskip

\item{} {\sl Technicolor (more properly, extended technicolor)
tends to have large flavor--changing neutral currents (FCNC for short). The
strange contortion of walking technicolor must be invoked to ameliorate
them. Supersymmetry does not have large FCNC.}

\medskip

\noindent Massive ETC boson exchange generically induces dangerously large
FCNC in the light quark and lepton sector unless $\Lambda_{ETC} \simge
500\,\tev$~\eekletc. In models in which the TC dynamics are QCD--like,
i.e., in which the coupling $\atc$ rapidly becomes small above
$\Lambda_{TC}$, this leads to estimates of quark and lepton masses from
Eq.~\qmass\ that are much too small for realism. In the ``walking
technicolor'' solution to the FCNC problem
\ref\wtc{B.~Holdom, Phys.~Rev.~{\bf D24} (1981) 1441;
{\it Phys.~Lett.}~{\bf 150B} (1985) 301\semi
T.~Appelquist, D.~Karabali and L.~C.~R. Wijewardhana,
Phys.~Rev.~Lett.~{\bf 57} (1986) 957\semi
T.~Appelquist and L.~C.~R.~Wijewardhana, Phys.~Rev.~{\bf D36} (1987) 568\semi
K.~Yamawaki, M.~Bando and K.~Matumoto, Phys.~Rev.~Lett.~{\bf 56}
(1986) 1335\semi
T.~Akiba and T.~Yanagida, Phys.~Lett.~{\bf 169B} (1986) 432.},
$\atc$ evolves very slowly with energy, remaining sizable for
a large energy range above $\Lambda_{TC}$. This implies an enhancement of
the right side of Eq.~\qmass\ which is of order
$\Lambda_{ETC}/\Lambda_{TC}$. This permits $\Lambda_{ETC}$ to be large
enough for the light quarks and leptons to suppress their FCNC interactions
to acceptable levels. Thus, the FCNC target was taken down long ago. I
don't know why people still shoot at it.

Some people find the unfamiliar dynamics of walking technicolor
``strange''; it certainly is very unlike QCD, in which the coupling quickly
starts to run above $1\,\gev$. I find this new dynamics mentally
liberating. It is stimulating---and humbling---to try to guess the spectrum
and other properties of a theory for which we have little theoretical and
no experimental input.

At the same time, it is misleading to say that supersymmetry has no FCNC
problem~\haber. For SUSY to avoid large FCNC, it is necessary that the mass
matrices of squarks and their corresponding quarks be simultaneously
diagonalizable, or almost so. In fact, since squark masses arise mainly
from soft SUSY--breaking terms, squarks of a given electric charge must be
nearly degenerate. To achieve this in supergravity models, it is assumed
that all soft SUSY--breaking terms renormalized at $M_P$ are universal.
Whether or not this assumption is natural is model--dependent. But, even if
it is natural, it is only in the technical ``set it and forget it'' sense
of supersymmetry.

\bigskip

\item{} {\sl Technicolor theories are in conflict with precision
electroweak measurements. Supersymmetry is not.}

\medskip

\noindent The effects of new physics, including technicolor, on
precisely--measured electroweak quantities have been studied for a long
time
\ref\pettests{
A.~Longhitano, Phys.~Rev.~{\bf D22} (1980) 1166; Nucl.~Phys.~{\bf B188},
(1981) 118\semi R.~Renken and M.~Peskin, Nucl.~Phys.~{\bf B211}
(1983) 93\semi
B.~W.~Lynn, M.~E.~Peskin and R~.G.~Stuart, in Trieste Electroweak 1985,
213\semi
M.~Golden and L.Randall, Nucl.~Phys.~{\bf B361} (1990) 3\semi B.~Holdom and
J.~Terning, Phys.~Lett.~{\bf 247B} (1990) 88\semi
M.~E.~Peskin and T.~Takeuchi, Phys.~Rev.~Lett.~{\bf 65} (1990) 964\semi
A.~Dobado, D.~Espriu and M.~J.~Herrero, Phys.~Lett.~{\bf 255B}
(1990) 405\semi
H.~Georgi, Nucl.~Phys.~{\bf B363} (1991) 301.}.
The so--called Peskin--Takeuchi parameter $S$, recently redetermined to be
$-0.15 \pm 0.25^{-0.08}_{+0.17}$~\pgl,
is claimed to be of $\CO(1)$ in most interesting TC models.

This is another case of shooting at the wrong target. The estimate $S_{TC}
\simge 1$ is known to be accurate only for those models with QCD--like
dynamics, i.e., in which a walking $\atc$ does {\it not} occur. In such
models, one may scale the relevant integrals from QCD (or actual hadronic
measurements). But, these QCD--like models were ruled out ages ago because
of their large FCNC. In walking technicolor models, the assumptions permitting
scaling from QCD are invalid, so that, in the absence of experimental input,
nonperturbative quantities such as $S$ cannot be reliably estimated
\ref\klichep{K.~Lane, {\it Technicolor and Precision Tests of the
Electroweak Interactions}, to appear in the Proceedings of the 27th
International Conference on High Energy Physics, Glasgow, 20--27th~July
1994; Boston University Preprint BUHEP--94--24 (1994).}.
There have been attempts to calculate $S$ in walking technicolor models that
have led to small and even negative values
\ref\tajtstu{B.~Holdom, Phys.~Lett.~{\bf 259B} (1991) 329\semi
E.~Gates and J.~Terning, Phys.~Rev.~Lett.~{\bf 67} (1991) 1840\semi
M.~Luty and R.~Sundrum, Phys.~Rev.~Lett.~{\bf 70} (1993) 127\semi
T.~Appelquist and J.~Terning, Phys.~Lett.~{\bf 315B} (1993) 139.}.
I think it is fair to say that, while these calculations suggest
TC models can produce an acceptable value of $S$, we still do not know how
to calculate intrinsically nonperturbative quantities in walking
technicolor.

\bigskip

\item{} {\sl It is difficult for technicolor to explain the large
top--quark mass.}

\medskip

\noindent This is the right target! Using Eq.~\qmass, the ETC scale
required to produce a top--quark mass of $175\,\gev$ is $\Lambda_{ETC}(t)
\sim 1\,\tev$. Such an ETC scale makes no sense dynamically because it is
too close to $\Lambda_{TC}$. To maintain a substantial hierarchy between
$\Lambda_{ETC}(t)$ and $\Lambda_{TC}$, it seems necessary that some ETC
interactions be strong enough to participate with TC in the breakdown of
electroweak symmetry {\it and} that the ETC coupling be fine-tuned (to
roughly a part in $10^2$--$10^6$, depending on details)
\ref\setc{T.~Appelquist, M.~B.~Einhorn, T.~Takeuchi and
L.~C.~R.~Wijewardhana, Phys.~Lett. {\bf 220B} (1989) 223\semi
T.~Takeuchi, Phys.~Rev.~{\bf D40} (1989) 2697\semi
V.A.~Miransky and K.~Yamawaki, Mod.~Phys.~Lett.~{\bf A4} (1989) 129\semi
V.~A.~Miransky, M.~Tanabashi and K.~Yamawaki, Phys.~Lett.~{\bf
221B} (1989) 177; Mod.~Phys.~Lett.~{\bf A4} (1989) 1043\semi
K.~Matumoto, Prog.~Theor.~Phys.~{\bf 81} (1989) 277.}.
But fine--tuning is the very demon technicolor was invented to exorcise.

\noindent Guilty as charged! In my opinion, understanding the top--quark
mass is the biggest challenge facing TC/ETC today. It is a difficult
problem. It is, in fact, one of the more glaring aspects of flavor
physics---the problem that remains unsolved at {\it all} energy scales by
{\it all} attempts to go beyond the standard model, including
supersymmetric ones.

\bigskip

\item{} {\sl Spontaneous symetry breaking implies a large
cosmological constant($\Lambda_0$), so TC/ETC, which has spontaneous
breaking at least up to 100s of TeV, has a terrible cosmological constant
problem.}

\medskip

\noindent This is also true and it is interesting to read 't~Hooft's
comments in his talk at this conference on how the cosmological constant
entered early considerations of spontaneously broken gauge symmetry. But
{\it all} theories with spontaneous symmetry breaking---QCD, the standard
Higgs model, the supersymmetric standard model, superstrings---have this
problem. And in gravity theories, one expects $\Lambda_0 \sim M_P^4$,
whereas we can infer the bound $\Lambda_0 \simle 10^{-121}  M_P^4$. As
Haber says~\haber, this is the ``mother of all fine--tuning and naturalness
problems.'' So, I'm not going to worry about it now.

\bigskip

\item{} {\sl Technicolor and extended technicolor are nonperturbative
theories, difficult to calculate. Supersymmetry is perturbative, easy to
calculate.}

\medskip

\noindent True: TC and, perhaps, ETC are strongly interacting field
theories. A walking gauge theory, if one exists, remains strong over a
broad range of energies and is about as nonperturbative an entity as I care
to contemplate. But, perturbation theory is not a law of nature: QCD is
also a nonperturbative field theory. Supersymmetry breaking may well be a
nonperturbative phenomenon~\haber. And, at the end of the day, i.e., at the
Planck scale, so are gravity, supergravity and superstring interactions. In
superstring theory---the Theory Of Everything---the mysteries of flavor lie
at $M_P$ where they are shrouded in that theory's nonperturbative delights.

\bigskip

\item{} {\sl Technicolor and extended technicolor are ugly---no
elegant model has been produced. Supersymmetry is beautiful.}

\medskip

\noindent {\it Nolo contendere.} But when I hear such statements {\it
without} supporting experimental evidence, I think of Boltzmann's remark
about elegance: It's for tailors and bootmakers.

\newsec{CLOSING REMARKS}

Technicolor plus extended technicolor is the most ambitious attempt yet to
explain the physics of electroweak and flavor symmetry breaking and to do
so in natural, dynamical terms. The effort has been frustratingly
difficult. This does not dissuade me and others from the TC/ETC philosophy
that the origin of this physics is to be found at energies far below the
Planck scale. Still, something important {\it is} missing. With the passage
of time, it seems inescapable that whatever that something is---possibly
including the answer that the whole ``low energy'' approach is wrong---will
be provided only by experiment.

Our best hopes for an answer in the ``near'' term lie with the Tevatron
Collider (where the top quark, perhaps our strongest hint on flavor
physics, seems to have been found) and with CERN's Large Hadron Collider.
Because of their higher and broader energy coverage, higher luminosity, and
more diverse beams, hadron colliders are the ideal tool for searching for
new physics when the details, especially the precise energy scale, are
unknown. However, I can tell you from close observation of real and
simulated hadron collider experiments, that those experiments are so
complicated that they are unlikely to find anything they are {\it not}
looking for. It is imperative, therefore, that their planning and execution
remain open to {\it all plausible} extensions of the standard model.

After all these years, we do not know what breaks electroweak symmetry,
what flavor symmetry means, and what breaks it. We are almost as far from
answers as Rabi and Glashow were over 30~years ago. We must keep our minds
open while we keep our experimental capabilities broad and unbiased. We
can't all believe in the same thing {\it until} experiments force us to. At
bottom, it is this need for us to remain open to what is really going on,
as well as my belief in electroweak and flavor physics at accessible energy
scales, that explains why I believe in technicolor.

\bigskip

I am indebted to the conference directors, Harvey Newman and Tom
Ypsilantis, for the privilege of attending this conference and hearing so
many wonderful, inspiring talks and to the Director, Nino Zichichi, and
staff of the Ettore Majorana Center for the magnificent organization and
execution of this conference and for the hospitality of the lovely town of
Erice. I thank Sekhar Chivukula, Howard Georgi, Mitchell Golden, Ryan Rohm,
Elizabeth Simmons and John Terning for their comments on the manuscript.
And I once again acknowledge the long collaboration and friendship of Tom
Appelquist and Estia Eichten. This work was supported in part by the
U.~S.~Department of Energy under Grant~No.~DE--FG02--91ER40676.

\listrefs
\vfil\eject
\bye